# Low Excess Noise, High Quantum Efficiency Avalanche Photodiodes for Beyond 2 μm Wavelength Detection

H. Jung[1†], S. Lee[1†], X. Jin[2†], Y. Liu[2], T. J. Ronningen[1], C. H. Grein[3], J. P. R. David[2], and S. Krishna[1*]

[1]Department of Electrical and Computer Engineering, The Ohio State University, Columbus, Ohio, 43210, USA
[2]Department of Electronic and Electrical Engineering, University of Sheffield, Sheffield S1 3JD, UK
[3]Department of Physics, University of Illinois, Chicago, Illinois 60607, USA
[*]Corresponding author: Sanjay Krishna, krishna.53@osu.edu
[†]These authors contributed equally to this paper

**Abstract**
The increasing concentration of greenhouse gases, notably $CH_4$ and $CO_2$, has fueled global temperature increases, intensifying concerns regarding the prevailing climate crisis. Effectively monitoring these gases demands a detector spanning the extended short-wavelength infrared (~2.4 μm) range, covering wavelengths of $CH_4$ (1.65 μm) and $CO_2$ (2.05 μm). The state-of-the-art HgCdTe avalanche photodetectors (APDs) offer exceptional performance metrics, including high gain ($M$) and low excess noise ($F$). However, their widespread adoption is hindered by inherent challenges such as manufacturability, reproducibility, and cost factors. Moreover, their reliance on cryogenic cooling adds to the cost, size, weight, and power of the system. We have demonstrated a linear mode APD combining an InGaAs/GaAsSb type-II superlattice absorber and an AlGaAsSb multiplier lattice matched to InP substrates. This APD has demonstrated a room temperature $M$ of 178, a maximum measurable external quantum efficiency of 3560 % at 2 μm, an extremely low excess noise ($F < 2$ at $M < 20$), and a small temperature coefficient of breakdown (7.58 mV/K·μm). Such a high performance APD with manufacturable semiconductor materials could lead to a rapid transition to a commercial III-V foundry, holding the promise of revolutionizing high-sensitivity receivers for greenhouse gas monitoring.

The concentration of greenhouse gases such as $CH_4$ and $CO_2$ has continuously risen in the last few decades, resulting in increased global temperatures, and raising concerns about a global climate crisis. Integrated path differential absorption (IPDA) and differential absorption LIDAR (DIAL) methods have been used as standard techniques for the measurement of trace atmospheric constituents. These methods are widely used in both ground-based[1,2] and airborne[3] research, establishing a foundation for potential future spaceborne missions[4]. To monitor greenhouse gas concentrations, it is, therefore, essential to have a detector that covers the extended short-wavelength infrared (eSWIR, ~ 2.6 μm) range since $CH_4$ (1.65 μm)[5] and $CO_2$ (2.05 μm)[6-8] have absorption features associated with vibration of the C-H and C-C stretch in this wavelength range. However, conventional *p-i-n* detectors are not suitable due to the reduction in signals received over long distances and harsh environments. Instead, avalanche photodiodes (APDs) are an attractive option because of the higher sensitivity provided by their internal multiplication gain, $M$. However, APDs have a stochastic gain mechanism that gives rise to excess noise simultaneously with signal multiplication, which poses a challenge in improving the signal-to-noise ratio. McIntyre's local field theory quantifies the excess noise factor, $F(M)$, as $F(M) = kM + (1-k)(2 - 1/M)$, where $k$ is the ratio of impact ionization coefficients ($\beta/\alpha$ for electron APDs and $\alpha/\beta$ for hole APDs)[9]. Thus, the selection of a material with a small impact ionization coefficient ratio is crucial to reduce $F(M)$.

Currently, HgCdTe (MCT) is the material of choice for high performance eSWIR and mid-wave infrared (MWIR) APDs because it provides high $M \sim 6100$ and extremely low $F(M) \sim 1.3$[10]. However, its high dark current from the narrow band gap necessitates a cooling system for efficient operation (at 77 – 110 K)[10]. Additionally, producing MCT with high uniformity is challenging, and the manufacturing process has low yield leading to increased cost. As such, researchers are seeking alternative materials. InAs is an emerging technology for SWIR APDs that offers a low $F(M) \sim 1.6$[11]. However, the narrow bandgap of the InAs also leads to a high tunneling dark current and the Fermi-level pinning in InAs increases the surface leakage current making it challenging to operate at temperatures above 77 K[11].

In general, for the eSWIR range, traditional *p-i-n* structure APDs are subject to the dominance of dark currents from the band-to-band tunneling mechanism, resulting from the narrow energy bandgap and high electric field, thereby making signal amplification difficult. To address this challenge, a separate absorption, charge, and multiplication (SACM) structure can be used to spatially isolate the absorption and multiplication regions. The SACM structure effectively suppresses the band-to-band tunneling mechanism, thereby enhancing the APD's ability to maintain a high electric field only in the wider gap multiplication region while keeping the electric field in the narrower gap absorber lower than its tunneling threshold. As a result, the SACM structure offers a significant advantage by reducing tunneling-induced noise while maintaining a high gain.

In 2020, Jones et al. demonstrated an $Al_xIn_{1-x}As_ySb_{1-y}$ (AlInAsSb)-based SACM APD on a GaSb substrate, detecting 2 μm wavelength[12]. This SACM APD presented a low excess noise (corresponding to $k \sim 0.01$) and low dark current density ($J_d \sim 33$ mA/cm at $M=10$) with an external quantum efficiency (EQE) of 20 % at 2 μm (at a punch-through, $V_{PT}$). Although the results are encouraging, the use of expensive GaSb substrates and more importantly, the complicated digital growth technique may hamper widespread adoption. To circumvent these limitations, we have been exploring an antimonide based SACM structure lattice matched to an InP substrate which is a commercial platform because of their affordability in a large area and low cost. There is a well-developed InGaAs/GaAsSb type-II superlattice (T2SL) absorber on InP[13,14], whose absorption spectrum can be extended to the eSWIR range. This absorber also can readily be grown because of the relatively thick layers of each T2SL constituent (5 nm) with each layer being lattice matched to InP. In 2006, Sidhu et al.[15] reported a SACM APD with a 5 nm-InGaAs/5 nm-GaAsSb T2SL absorber with an InP multiplier. The device provided $M > 30$ at room temperature and a cut-off wavelength of 2.4 µm. However, its performance was limited by the high excess noise of the InP ($k \sim 0.5$) multiplication region. A few years later, Goh et al.[16] and Ong et al.[17] demonstrated APDs with InGaAs/GaAsSb T2SL absorbers with a slightly lower excess noise InAlAs multiplier. These devices demonstrated EQE of about 20% at 2 µm operating at room temperature. Although the excess noise was reduced by replacing InP with InAlAs ($k \sim 0.2$), it still requires improvement to meet the demands of the current application and to be comparable to the performance of the AlInAsSb SACM APDs on GaSb substrates.

$Al_xGa_{1-x}As_ySb_{1-y}$ is a quaternary alloy that can be grown lattice-matched on InP. Initial work by Yi et al.[18,19] showed that $AlAs_{0.56}Sb_{0.44}$ (AlAsSb) can give rise to a very large $\alpha/\beta$ ratio in thick bulk structures and a very low $F(M)$ of ~2.1 and $M$~36. More recently, Lee et al.[20] have explored $Al_{0.85}Ga_{0.15}AsSb$ (AlGaAsSb) with a 1000 nm-thick unintentionally doped layer grown as a random alloy, lattice-matched to the InP substrate. Their findings revealed that this material has a remarkably low excess noise corresponding to $k \sim 0.01$, with a gain of up to 278 achievable under 1550 nm illumination while operating at room temperature. Consequently, it has emerged as a promising material candidate for next-generation APD technology. In this paper, we have combined the AlGaAsSb multiplier with an InGaAs/GaAsSb T2SL absorber, both lattice-matched to an InP substrate, for the first time to achieve high-performance SACM APDs for eSWIR detection (cutoff ~2.4 µm). Our device exhibits high gain-quantum efficiency product (3560 % at 2 µm), low excess noise ($F < 2$ at $M < 20$), and high gain ($M \sim 178$) at room temperature, providing advanced eSWIR SACM APD performance on an InP substrate. These results demonstrate the potential of this material combination for use in eSWIR detection.

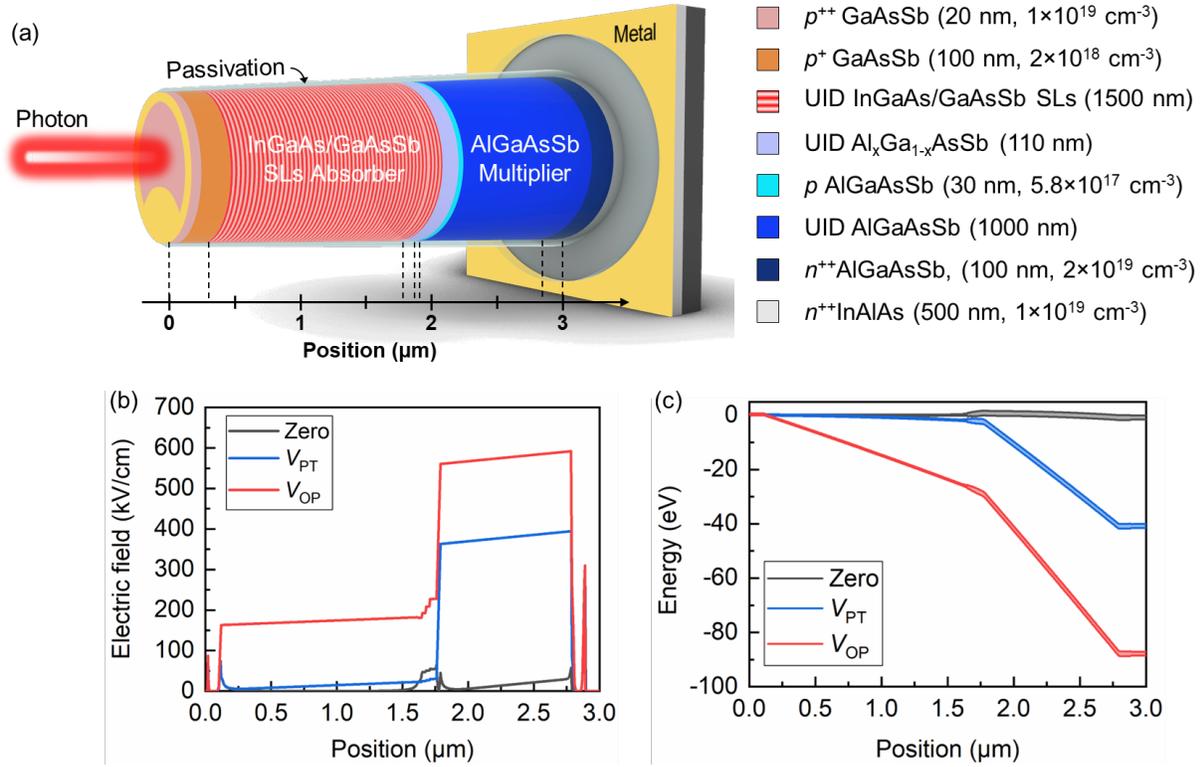

Fig. 1 (a) Schematic eSWIR SACM APD device structure with InGaAs/GaAsSb T2SL absorber and AlGaAsSb multiplier. (b) Simulated electric fields, and (c) band profiles of the eSWIR SACM APD with a nominal thickness at zero, punch-through ($V_{PT}$), and operating ($V_{OP}$) bias.

**Results**

**Device design and simulation.** The high-performance SACM APDs designed for eSWIR detection require high material quality with low impurity concentrations. The unintentionally doped (UID) layer thicknesses for the absorber and the multiplier were designed to be 1500 nm and 1000 nm, respectively, with the specific structural details as illustrated in Fig. 1 (a). The $Al_xGa_{1-x}As_ySb_{1-y}$ grading layer between the absorber and the multiplier is composed of three different Al compositions: 0, 25, and 55%. The AlGaAsSb quaternary material in this SACM APD was grown as a random alloy, and it utilized a InGaAs/GaAsSb T2SL absorber lattice matched to the InP substrate, both of which can be provided by a commercial foundry. Details of the material growth and X-ray diffraction (XRD) results of this device are provided in the *Methods* and the *Supplementary Information* (*S1*), respectively. The simulated electric fields and energy band profiles of the device at zero, $V_{PT}$, and operational ($V_{OP}$) bias after punch-through are illustrated in Fig. 1(b) and Fig.1 (c), respectively, and were obtained using Silvaco software. Here, $V_{OP}$ is -87 V, where avalanche multiplication occurs. A well-designed grading layer between the absorber and multiplier allows for smooth electron transport over the operational bias range.

**Capacitance-voltage, current-voltage, and gain.** Capacitance-voltage (*C-V*) measurements were carried out to identify the doping profile and the depletion width of an InGaAs/GaAsSb T2SL-based SACM APD, as shown in Fig. 2 (a). A sudden drop in capacitance around -48 V indicates the occurrence of punch-through. The simulated *C-V* curve by a 1D Poisson solver closely matches the measured $V_{PT}$, as indicated by the dotted black line. The inset in Fig 2 (a) shows that the Be dopants appear to have undergone some diffusion, resulting in a secondary peak at the start of the T2SL region. The integrated charge is only 2.8% higher than the design value. The total thickness of the InGaAs/GaAsSb T2SL absorber was modeled as 1400 nm, similar to values obtained from XRD (see *S1* in *Supplementary Information*). Both the absorber and multiplier have background doping concentrations of $1 \times 10^{15}$ and $4 \times 10^{15}$ cm$^{-3}$, respectively. Fig. 2 (b) shows the dark current density of devices with 150, 200, 250, and 300 μm diameters. The dark current densities increase sharply at -48 V, consistent with the $V_{PT}$ obtained from the *C-V* curve. The dark current densities do not scale with the area before punch-through, however are overlapped after punch-through, suggesting that the dark currents of the multiplier and absorber were limited by the surface component and the bulk component, respectively. The gain curve in Fig. 2 (c) was obtained under 940 nm wavelength illumination at room temperature, and the dotted black line is a fitted curve simulated by using the Random Path Length (RPL) model[21]. The maximum gain our SACM APD achieved was about 178, but this value was limited by edge breakdown rather than the inherent characteristics of the devices. A well-designed planar device architecture should provide a much higher gain value.

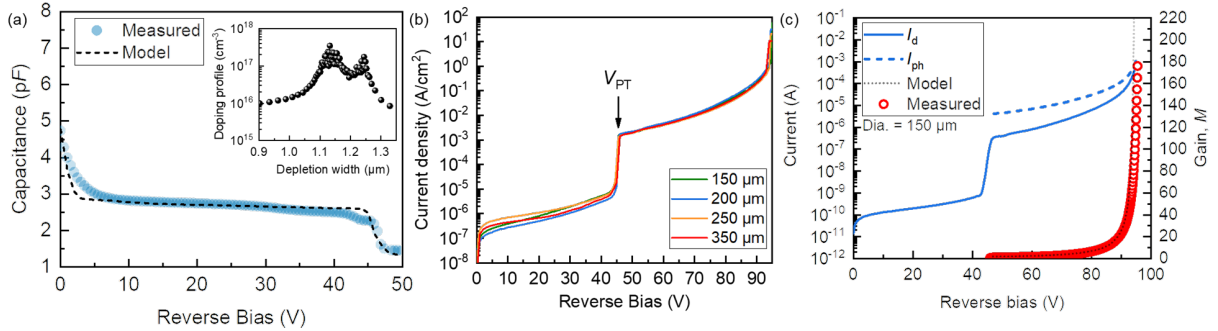

Fig. 2 (a) Measured (blue dot) and simulated (black-dashed line) capacitance-voltage results. Inset in (a) shows the doping profile of the charge sheet layer. (b) Dark current density from various sized devices as a function of reverse bias. The punch-through ($V_{PT}$) happens at around 48 V in both *C-V* and *I-V*. (c) shows photocurrent and gain from 150 µm device obtained with 940 nm illumination. The measured gain agrees well with the fitted curve modeled using impact ionization coefficients of the AlGaAsSb.

**External quantum efficiency and excess noise.** Spectral response measurements were used to determine the EQE of the extended SACM APDs. Fig. 3 (a) displays the spectral EQE obtained from the relative responsivities of a SACM APD under various voltages at room temperature. The device has a cutoff wavelength of ~2.4 µm, and the EQE at 2 µm was 20 % at -48V and remained constant until -50 V, indicating that the punch-through at -48 V is consistent with the values obtained from *C-V* and *I-V*. The device can achieve a gain×QE product up to 3560 % at 2 µm, and the multiplication calculated from the EQE results at 2 µm compares very well with the gain obtained by the photocurrent, as shown in Fig. 3 (b).

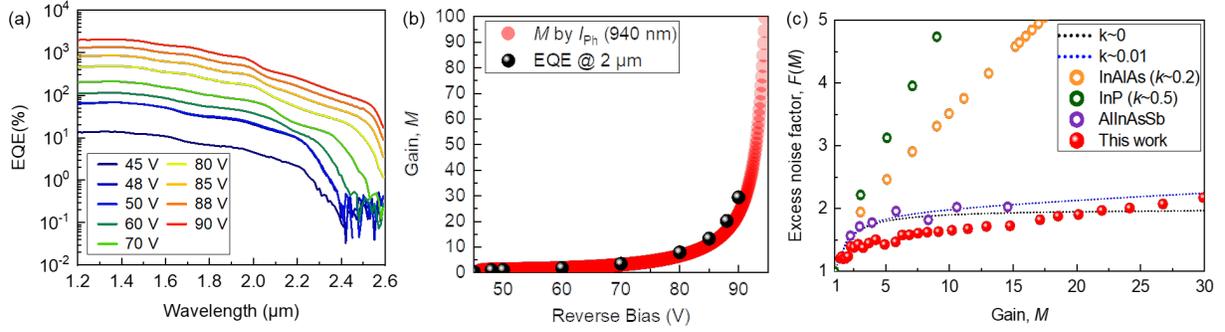

Fig. 3 (a) EQE of eSWIR SACM APD at various applied biases, and (b) extracted gain from EQE at 2 μm compared to the gain obtained by photocurrent under 940 nm laser illumination as a function of reverse bias. (c) Measured excess noise as a function of gain with other emerging III-V-based SACM APDs[12,15,17] and theoretical values based on the McIntyre local field model when $k\sim0$ (black-dashed line) and $k \sim 0.01$ (blue-dashed line).

Fig. 3 (c) shows the measured $F(M)$ of this SACM APD using the setup as detailed in the Methods section. Notably, our device exhibits an exceptionally low $F(M)$, nearing $F(M) \sim 1.7$ at $M = 10$. This value is approximately two times lower than that achieved in previous research involving the InGaAs/GaAsSb T2SL with an InAlAs multiplier on InP[17]. Furthermore, our T2SL SACM APD can have a 2.5 times higher operating gain value ($M = 25$) for the equivalent $F(M=10)$ of AlInAsSb APDs grown on InP. For $M$ less than 20, the only APD that has exhibited noise performance superior to our device is a MCT APD[10,22]. However, it is crucial to note that the MCT requires operation at considerably lower temperatures, typically within the range of 77 to 110 K[10,22,23], thereby presenting challenges in terms of high cost, size, weight, and power (c-SWaP). Additionally, the $F(M)$ of this work is even lower than McIntyre's local field model limit at $M < 20$, possibly due to the non-exponential Weibull-Fréchet probability distribution function of the impact ionization process[24].

**Temperature coefficient of breakdown.** One of the essential performance metrics in APDs is the temperature coefficient of breakdown ($C_{bd}$), which is defined as the rate of change in the breakdown voltage with temperature. Normally, the breakdown voltage of an APD tends to increase as the temperature rises, primarily due to enhanced phonon scattering reducing the ionization coefficients. This phenomenon can have a significant impact on the gain, and consequently, the sensitivity of APDs made from materials such as InP and Si, where $C_{bd}$ is large.

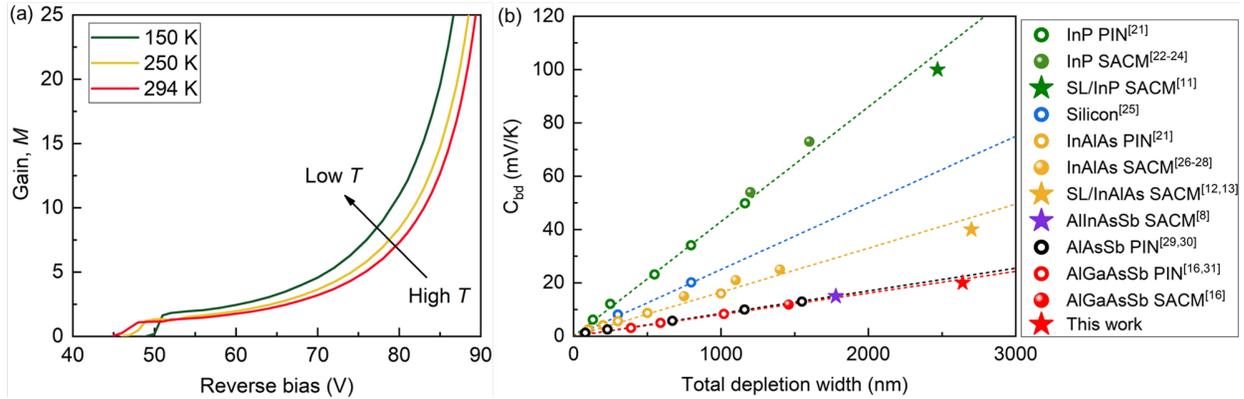

Fig. 4 (a) Measured gain, $M$, as a function of reverse bias at three different temperatures: 150, 250, and 294 K. (b) $C_{bd}$ versus total depletion width for APDs of various materials: InP[15,25-28], Si[29], InAlAs[16,17,25,30-32], AlInAsSb[12], AlAsSb[33,34], and AlGaAsSb[20,35]. Star symbol refers to devices operating at $\geq 2$ μm wavelength.

Any change in breakdown voltage necessitates the incorporation of temperature stabilization in a LiDAR system. Therefore, it is advantageous to have a low $C_{bd}$, especially since bulky temperature control systems are not favored in spaceborne or airborne lidar systems, which aim to minimize their size, weight, and power consumption.

To attain the $C_{bd}$ of our device, gain measurements were conducted at different temperatures: 150, 250, and 294 K, as depicted in Fig. 4. The assessment of the voltage change with temperature was performed at $M = 25$ rather than at the actual breakdown point. This approach was chosen to yield similar results while minimizing the risk of catastrophic damage to the devices. The accurate determination of the device junction temperatures was carried out using the methodology outlined elsewhere[34]. The measured $C_{bd}$ is approximately 20 mV/K for the total thickness of 2640 nm, which is approximately 4.75 and 2 times lower than the value observed for the 2470 nm thick T2SL/InP SACM APD (100 mV/K)[15], the 2755 nm T2SL/InAlAs SACM APD (~ 40 mV/K)[16,17], respectively, and was similar to a 1780 nm thick AlInAsSb SACM APD (~18 mV/K)[12]. However, it should be noted that $C_{bd}$ is a function of the total device width (see Fig. 4b). When we normalize our $C_{bd}$ by the total thickness in µm ($C_{bd}$ /µm), it is apparent that our SACM APD has a $C_{bd}$ /µm of 7.58 mV/K·µm which is smaller than 8.43 mV/K·µm for AlInAsSb SACM APD. The $C_{bd}$ of our SACM APD is, therefore, extremely low compared to other T2SL SACMs, even though their total thicknesses are similar.

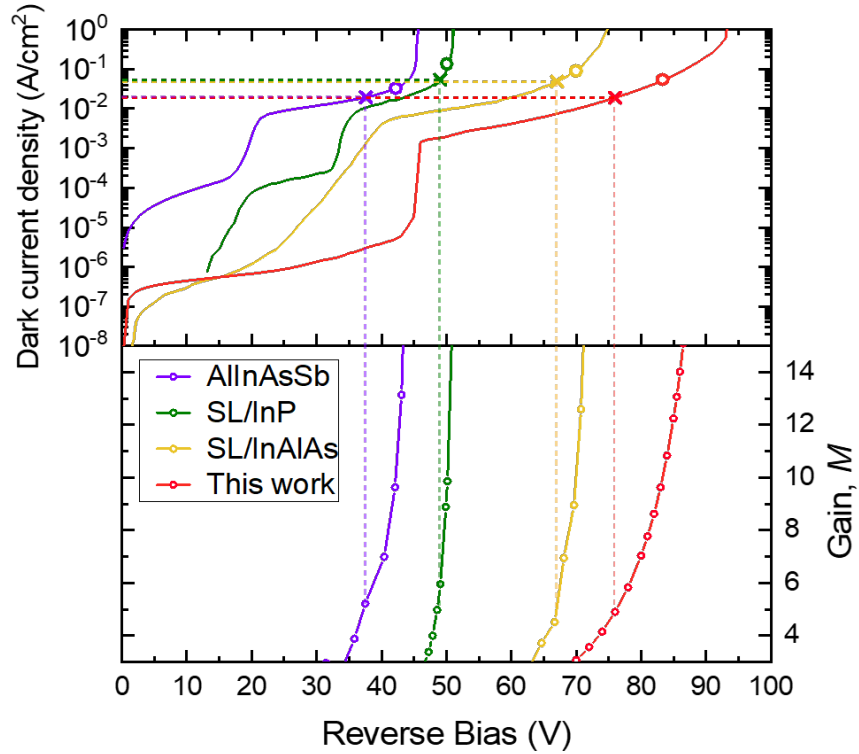

Fig. 5 Dark current density (upper) and gain (lower) vs. reverse bias of this work (red) compared with other emerging III-V eSWIR SACM APDs: AlInAsSb (purple)[12], InGaAs/GaAsSb T2SL with InP (green)[15] and InGaAs/GaAsSb with InAlAs (yellow)[16]. The cross and circular marks on each dark current density curve indicate the dark current density where $M=5$ and $M=10$, respectively.

**Discussion**

The dark current density of our T2SL SACM APD is compared with other emerging III-V eSWIR SACM APDs in Fig. 5. In particular, the dark current density after $V_{pt}$ is of interest since shot noise of a SACM APD is proportional to the square root of the dark current at the operating bias. After $V_{pt}$, our device shows the dark current density that is one order of magnitude lower than other APDs. As the applied reverse voltage increases, the dark current density of our device rapidly increases and finally becomes comparable to others. However, it is important to highlight that this elevated dark current density can be effectively reduced through careful device design optimization, for example, by increasing the charge sheet doping or thickness so that the electric field in the T2SL absorber is reduced while the field in the multiplication region increases This design would not only reduce the dark current in the absorber while enhancing the gain, but would also reduce the operating bias as well.

Table 1 presents a comparative analysis of our work alongside other emerging III-V-based eSWIR APDs, including an AlInAsSb SACM APD on GaSb, and an InGaAs/GaAsSb T2SL SACM APD with InP and InAlAs multipliers. Our AlGaAsSb-based SACM APD, incorporating an InGaAs/GaAsSb T2SL, stands out for its exceptional performance in all metrics. Especially, it achieves the lowest $F(M)$ at $M \sim 10$ and exhibits the lowest dark current density at $V_{PT}$ and even at $M\sim 5$. While our SACM APD shows a slightly higher dark current density at $M \sim 10$ (55 A/cm$^2$) when compared to the AlInAsSb SACM APD device B (33 mA/cm$^2$) it should be noted that our device features a cut-off at 2.4 μm (0.516 eV), in contrast to the 2.1 μm (0.59 eV) cut-off of AlInAsSb, which constitutes a significant energy difference of 0.074 eV. To compare these two materials more fairly, one should refer to the IGA-17 Rule as depicted in the *Supplementary Information (see Fig. S2)*, which shows that the dark current of the results shown here relative to the bandgap is actually very low. Temperature dependence of the dark currents from 300 K to 240 K showed an activation energy of 0.375 meV (see Fig. S3 and S4 in *Supplementary Information*) in good agreement with the activation energies reported for other T2SL structures[15]. The longer cutoff wavelength of 2.4 µm in our structure c.f. the AlInAsSb APD means that more cooling can be applied to reduce the dark currents without losing the ability to operate at 2 µm.

**Table 1** Comparison of this work with other emerging III-V eSWIR SACM APDs[12,15,16].

| Materials | AlInAsSb (Device B) | InGaAsSb SL/ InP | InGaAsSb SL/InAlAs | This work |
|---|---|---|---|---|
| Substrate | GaSb | InP | InP | InP |
| Max reported $M$ | 150 | 35 | 60 | 178 |
| $J_d$ @ $V_{PT}$ | 8 mA/cm² | 8.55 mA/cm² ($M$~1.7) | 5 mA/cm² | 1.9 mA/cm² |
| $J_d$ @ $M = 5$ | 19 mA/cm² | 54 mA/cm² | 48 mA/cm² | 18 mA/cm² |
| $J_d$ @ $M = 10$ | 33 mA/cm² | 136 mA/cm² | 90 mA/cm² | 55 mA/cm² |
| $F(M = 10)$ | 2.02 | 5.13 | 3.51 | 1.65 |
| Cut-off λ | 2.1 µm | 2.4 µm | 2.5 µm | 2.4 µm |
| $C_{bd}/W$ | 8.43 mV/K·µm | 40.49 mV/K·µm | 14.52 mV/K·µm | 7.58 mV/K·µm |

In this work, we have demonstrated SACM APDs with antimonide-based materials on InP substrates for efficient detection beyond 2 µm wavelength at room temperature. At 2 µm wavelength, our device exhibits a multiplied QE of 3560 % ($M = 178$) with extremely low excess noise ($F < 2$ at $M < 20$) at room temperature and a smallest temperature coefficient of breakdown ($C_{bd}$/µm = 7.58 mV/K·µm) among other III-V APD technologies on InP substrate. In addition to being grown on a standard InP substrate, the multiplication region and the T2SL absorber regions are grown using well-established growth techniques, meaning that these APDs can be easily manufactured.

**Acknowledgments**

This work was supported by the Advanced Component Technology (ACT) Program of NASA's Earth Science Technology Office (ESTO) under Grant No. 80NSSC21K0613.

**Author contributions**

H.J. and S.L. designed the structures and performed material growth, fabrication, and preliminary device characterizations for all InGaAs/GaAsSb superlattice/AlGaAsSb SACM structures. X.J. and Y.L. undertook IV, CV, QE, multiplication, and excess noise for the SACM structure. All authors discussed and analyzed the results, and H. J., S.L., X.J., Y.L., T.J.R., C.H.G., J.P.R.D., and S.K. wrote the manuscript. J.P.R.D helped with data analysis and S.K. supervised the entire project. All authors reviewed and approved the manuscript.

**Competing interests**

The authors declare no competing interests.

**Methods**

**Material Growth.** The eSWIR SACM APD was grown on *n*-InP substrates using a solid-state MBE reactor. For group V cells, we used RIBER VAC 500 and Veeco Mark V valved crackers for As and Sb, respectively. To achieve very low background doping concentration for both InGaAs/GaAsSb SL and random alloy AlGaAsSb, the calibration growths and several $p^+$-$i$-$n^+$ growths were performed at various growth conditions such as growth rate, V/III beam equivalent pressure (BEP) ratio, and growth temperature. More details on the growth can be found elsewhere[36-38].

**Device Fabrication.** To assess the performance of the device, the epi materials underwent fabrication using citric-based wet chemicals for mesa delineation. Subsequently, the device's sidewall was covered by SU8 to suppress the surface leakage current. Finally, Ti/Au (12/150 nm) was deposited for top and bottom *n*-contacts using an *e*-beam evaporator.

**I-V and C-V Measurements.** The electrical performances of the device were assessed by *I-V* and *C-V* characteristics. The *I-V* characteristics were measured using a probe station and an HP4140B picoammeter. *C-V* measurements were undertaken using an HP4275A LCR meter as a frequency of 10 KHz. To calculate the depletion width and background doping concentration of the device, a static dielectric constant of 11.4 was used for the AlGaAsSb multiplier, and a value of ~ 14 was taken as an average weight for InGaAs and GaAsSb.

**Multiplication and Excess Noise.** A transimpedance-amplifier-based circuit is employed to assess the multiplication and excess noise in these structures, featuring a center frequency of 10

MHz and bandwidth of 4.2 MHz as described in elsewhere[39]. To eliminate the effects of the DC leakage current, we utilized phase-sensitive detection. The shot noise was calibrated by using a reference device (SFH2701 Silicon PIN photodiode). To determine the excess noise factor of the AlGaAsSb-based eSWIR APD, the measured noise power of our device was compared to the measured noise power of the reference device at a given photocurrent. A Thorlabs fiber-coupled LED (M1450F1) was used to illuminate 1450 nm on the devices to measure the multiplication and the excess noise. The unity gain value of the T2SL/AlGaAsSb SACM was determined by fitting the multiplication curve to the random path length model.

**Quantum Efficiency.** QE measurements involved illuminating a 100 W tungsten bulb into a monochromator (IHR320), which was focused onto the device under test using optical lenses, while the photocurrent was measured using a lock-in amplifier. A mechanical chopper modulated the light at 180 Hz to remove DC dark current, and a Keithley 236 source meter unit was used to apply the bias. As a reference sample, a commercial extended InGaAs photodiode (Thorlabs FD05D) was used to determine the relative power of the monochromator at each wavelength with known responsivity.

**Data availability**

The data that support the plots within this paper and other finding of this study are available from the corresponding authors upon reasonable request.